\documentclass[dvips,12pt,a4paper]{article}
\usepackage{a4wide}
\usepackage{epsfig}
\usepackage{amsmath}
\usepackage{latexsym}

  \newlength{\absize}
\newcommand{\dd}{\mbox{{\rm d}}}

\renewcommand{\Re}{\mbox{Re}}

\newcommand{\Order}{{\cal O}}
\newcommand{\TeV}{{\rm TeV}}
\newcommand{\Lumint}{{\cal L}_{\rm int}}
\def \sup{^{\vphantom{2}}}
\begin{document}
  \thispagestyle{empty}
  \pagestyle{empty}
  \renewcommand{\thefootnote}{\fnsymbol{footnote}}
\newpage\normalsize
    \pagestyle{plain}
    \setlength{\baselineskip}{4ex}\par
    \setcounter{footnote}{0}
    \renewcommand{\thefootnote}{\arabic{footnote}}
\newcommand{\preprint}[1]{%
  \begin{flushright}
    \setlength{\baselineskip}{3ex} #1
  \end{flushright}}
\renewcommand{\title}[1]{%
  \begin{center}
    \LARGE #1
  \end{center}\par}
\renewcommand{\author}[1]{%
  \vspace{2ex}
  {\Large
   \begin{center}
     \setlength{\baselineskip}{3ex} #1 \par
   \end{center}}}
\renewcommand{\thanks}[1]{\footnote{#1}}
\renewcommand{\abstract}[1]{%
  \vspace{2ex}
  \normalsize
  \begin{center}
    \centerline{\bf Abstract}\par
    \vspace{2ex}
    \parbox{\absize}{#1\setlength{\baselineskip}{2.5ex}\par}
  \end{center}}
\preprint{University of Bergen, Department of Physics \\
Scientific/Technical Report No.\ 1996-14 \\ ISSN~0803-2696\\
December 1996}
\vfill
\title{$Z'$ interference effects from TRISTAN to LEP2}

\vfill
\author{P. Osland, A.A. Pankov\footnote{Permanent address: Gomel
Polytechnical Institute, Gomel, 246746 Belarus.} \\\hfil\\
        Department of Physics\thanks{Electronic mail addresses:
                {\tt Per.Osland@fi.uib.no; pankov@gpi.gomel.by}}\\
        University of Bergen \\ All\'egt.~55, N-5007 Bergen, Norway}
                                                                       
\vfill
\abstract
{We discuss the potential of present $e^+e^-$ colliders to probe for
a heavy $Z^\prime$ boson via its interference effects, 
in a  model-independent framework. 
The leptonic channel is unambiguous, and thus favourable over 
the hadronic ones.
The effect on the leptonic cross section $\sigma_{\mu\mu}$,
induced by a heavy $Z^\prime$ with arbitrary couplings, 
is uniquely determined at LEP2 energies.
It results in a {\it negative} deviation from the standard model prediction,
due to destructive $\gamma-Z'$ and $Z-Z'$ interferences.
At TRISTAN energies the former dominates and is given by the vector 
coupling of the $Z^\prime$.
Around the $Z$ peak, the $Z-Z'$ interference dominates 
and depends on the axial-vector coupling. 
The forward-backward asymmetry is complementary to 
$\sigma_{\mu\mu}$ in the $Z^\prime$ search. 
Comparison of the $e^+e^-$ potentials indicates that already after one 
year of operation, LEP2 will achieve a sensitivity to $Z^\prime$ 
that exceeds the current one at TRISTAN.}
                
\newpage
    \setcounter{footnote}{0}
    \renewcommand{\thefootnote}{\arabic{footnote}}
    \setcounter{page}{1}
\section{Introduction}
%%%%%%%%%%%%%%%%%%%%%%%%%%%%%%%%%%%%%%%%%%%%%%%%%%%%%%%%%%%%%%%%%%%%%%%%
Although the success of the standard $SU(2)_{\rm L}$$\times$$U(1)_{\rm Y}$
electroweak model (SM) is certainly unquestionable in describing 
the observed interactions of quarks and leptons, 
from the low energies involved in atomic physics 
parity violation experiments up to the high energies accessible at LEP 
and the Tevatron \cite{Langacker,Warsaw}, 
other models with one or more additional gauge bosons 
($Z'$) are not ruled out \cite{Rizzo}. A major objective of experiments at
present and future high-energy accelerators is to search for
new particles and interactions that would announce the onset of
physics beyond the SM. 

While open thresholds provide the most unambiguous signals 
of new physics, the mass range is in this case strictly limited
by the available beam energy.
In fact, at present $e^+e^-$ colliders, TRISTAN and LEP2, a
$Z'$ would not be directly produced, because of the already 
existing mass limits \cite{Godfrey,Maeshima}.
A considerably larger mass range may be explored, however, through
the study of ``virtual effects" \cite{lep2}.
For masses much larger than the beam energy ($M_{Z'}\gg \sqrt{s}$)
such indirect signatures are usually investigated within 
the frameworks of specific models.
The consistency of experimental data with the SM is usually 
interpreted in terms of $Z'$ mass limits \cite{Godfrey,Leike1} for specific
models (e.g., $E_6$, the left-right symmetric model, 
the left-right alternative model, or the sequential standard 
model \cite{Rizzo}). Such limits give a valuable feeling for the 
resolving powers of specific reactions and experiments, but the full 
information about a possible $Z'$ can only be obtained by a 
model-independent analysis of the data \cite{Leike2}.
In this regard, most attention has been given to the annihilation
processes
\begin{eqnarray}
\label{Eq:leptons}
e^+e^-&\to& l^+l^-, \qquad \mbox{($l=\mu$, $\tau$)}, \\
\label{Eq:quarks}
e^+e^-&\to& q \bar q.
\end{eqnarray}
The reaction (\ref{Eq:leptons}) which is very clean and with good 
statistics, is preferred for such a dedicated and model-independent analysis
because it involves only leptonic couplings of the $Z'$, in contrast to 
the process (\ref{Eq:quarks}).

As mentioned above, we have no experimental indications for extra neutral
gauge bosons. Nevertheless, a recently observed phenomenon at 
TRISTAN \cite{kek1,kek2} deviates from this picture. 
Indeed, the cross section for muon pair production is slightly lower 
than the SM expectation by two standard deviations \cite{kek2}.
A similar tendency has been observed at LEP1 in a study of
radiative muon-pair events at $\sqrt{s}<M_Z$ \cite{lep1-rad}.
If this discrepancy is real, it indicates a problem with the SM.

This modest deviation of $\sigma_{\mu\mu}$ finds 
a natural explanation in theories with an extended gauge sector,
with additional neutral vector bosons. The analysis of this possibility
in the framework of theories  based on the $E_6$ gauge group was performed 
in \cite{p1}. It was shown that for several models the deviation of 
$\sigma_{\mu\mu}$ from the SM prediction at TRISTAN energies can be explained 
by the destructive contribution of the $\gamma-Z'$ interference.  
It should be noted, however, that the sensitivity of TRISTAN is only
sufficient to observe the $\gamma-Z'$ interference effects in a leptonic 
channel if the upper limit on the additional $E_6$ gauge boson mass 
does not exceed 300--350~GeV. However, the
present limit on the $E_6$ $Z'$ boson mass \cite{Godfrey} obtained 
from direct searches at the Tevatron collider lies higher,
$M_{Z'}>\Order(0.5~\TeV)$ \cite{Maeshima}. 
This means that additional $Z'$ bosons with 
such high masses in the framework of theories based on the $E_6$ group 
have no chance of explaining the effect observed at TRISTAN. 

However, another possibility 
is the explanation \cite{p2} based on the additional $Z^\prime$ 
(referred to as isoscalar weak vector bosons $Y$ and $Y_L$)
arising in alternative models with composite structure of electroweak 
interactions \cite{kuroda}.
It is interesting to note that the possible existence of such
$Z^\prime$ bosons might lead to new interference effects \cite{p2} 
in $e^+e^-\to f\bar f$ at LEP1 energies. 
Namely, in addition to the $\gamma -Z^\prime$ interference term, the
contribution of the $Z-Z^\prime$ interference becomes significant at
energies $M_Z\mp\Gamma_Z/2$ and manifests itself as a specific
interference pattern (peak and dip).

Although the analysis above has been done within specific models,
and in a limited energy range $\sqrt s\le M_Z$, 
a more complete information on heavy neutral bosons can be obtained from 
a model-independent $Z'$ analysis of the data
available at present $e^+e^-$ colliders.
This model-independent approach allows, on the one hand, 
to cover a wide class of models, and, on the other hand 
to provide a comparison 
of the potentials of present $e^+e^-$ colliders to such $Z^\prime$ searches. 
In this paper we extend the previous Born-level analyses \cite{p1,p2} 
by taking into account all radiative corrections, 
and also consider a higher energy range, up to 190~GeV. 
We also discuss the above-mentioned deviation of the cross section 
$\sigma_{\mu\mu}$ from the SM prediction at TRISTAN, 
and its implication for LEP2.

Thus, the purpose of this paper is to present a comparative and
model-independent analysis of the potential to search for $Z'$
interference effects at present $e^+e^-$ colliders.
We concentrate on the muon pair production, which is particularly 
suitable for this purpose. The entire available energy range, 
from 58 GeV (TRISTAN) up to 190 GeV (LEP2), is included in this study, 
based on realistic assumptions about the expected detector performances 
and machine luminosities.
In addition, in this paper we have tried to answer two complementary 
questions.
The first question is: Which information on a $Z'$ can one derive 
if no indirect signal of any type is observed at present 
$e^+e^-$ colliders?
To answer this question one is led to the derivation of certain
confidence limits on the $Z'$ parameters. This will be done in a
model-independent manner. The second question is: If an indirect signal 
were seen at TRISTAN, would it be possible to observe it at LEP2 with 
sufficient significance that one could either confirm or reject 
the hypothesis?

%%%%%%%%%%%%%%%%%%%%%%%%%%%%%%%%%%%%%%%%%%%%%%%%%%%%%%%%%%%%%%%%%%%%%%%%
\section{Preliminaries}
%%%%%%%%%%%%%%%%%%%%%%%%%%%%%%%%%%%%%%%%%%%%%%%%%%%%%%%%%%%%%%%%%%%%%%%%
A new neutral gauge boson induces additional neutral current 
interactions, the corresponding Lagrangian can be written as
\begin{equation}
-{\cal L}_{\rm NC}=eJ^{\mu}_{\gamma}A_{\mu}+
g_ZJ^{\mu}_ZZ_{\mu}+g_{Z'}J^{\mu}_{Z'}Z'_{\mu},
\end{equation}
where $e=\sqrt{4\pi\alpha}$, $g_Z=e/s_Wc_W$ 
($s^2_W=1-c^2_W\equiv\sin^2\theta_W$) and $g_{Z'}$ are the gauge coupling
constants. The neutral currents are
\begin{equation}
J^\mu_i=\sum_{f}\bar\psi_f\gamma^\mu\left(L^f_iP_L+R^f_iP_R\right)\psi_f=
\sum_{f}\bar\psi_f\gamma^\mu\left(V^f_i-A^f_i\gamma_5\right)\psi_f,
\label{current}
\end{equation}
where $i\equiv\gamma$, $Z$, $Z'$, and $P_{L,R}=(1\mp\gamma_5)/2$ are the 
left- and right-handed chirality projection operators. The SM 
left- and right-handed couplings of the vector boson $i$ to the fermions 
are
\begin{equation}
L^f_\gamma=Q_f, \qquad R^f_\gamma=Q_f, \qquad
L^f_Z=I^f_{3L}-Q_fs_W^2, \qquad R^f_Z=-Q_fs_W^2,
\end{equation}
or, in terms of vector and axial-vector couplings:  
\begin{equation}
V^f_\gamma=Q_f, \qquad A^f_\gamma=0, \qquad
V^f_Z=\frac{I^f_{3L}}{2}-Q_fs_W^2, \qquad A^f_Z=\frac{I^f_{3L}}{2}.
\end{equation}
Here, $Q_f$ is the electric charge of $f$ ($Q_e=1$),
and $I^f_{3L}$ denotes the third component of the weak isospin.

The lowest-order unpolarized differential cross section for the processes
(\ref{Eq:leptons}) and (\ref{Eq:quarks}), including $\gamma$, 
$Z$, and the extra $Z'$ boson exchanges, is given by
\begin{eqnarray}
\frac{\dd\sigma_{ff}}{\dd\cos\theta}
&=&\frac{\pi\alpha^2}{2s}
\left[(1+\cos^2\theta)\, F_1 +2\cos\theta\, F_2\right], \nonumber \\
F_1&=&F_1^{\rm SM}+\Delta F_1, \qquad F_2=F_2^{\rm SM}+\Delta F_2,
\end{eqnarray}
with
\begin{eqnarray}
F_1^{\rm SM}&=&Q^2_eQ^2_f+2\,Q_ev_eQ_fv_f\Re\chi\sup_Z
+(v^2_e+a^2_e)(v^2_f+a^2_f)|\chi\sup_Z|^2,\nonumber \\[2mm]
F_2^{\rm SM}&=&2\,Q_ea_eQ_fa_f\Re\chi\sup_Z
+4\,v_ea_ev_fa_f|\chi\sup_Z|^2, \nonumber \\[2mm]
\Delta F_1&=&2\,Q_ev'_eQ_fv'_f\Re\chi\sup_{Z'}
+(v_e'{}^2+a_e'{}^2)(v_f'{}^2+a_f'{}^2)|\chi\sup_{Z'}|^2 \nonumber \\[2mm]
& &
+2\,(v_ev'_e+a_ea'_e)(v_fv'_f+a_fa'_f)\Re(\chi\sup_{Z}\chi^*_{Z'}),
\nonumber \\[2mm]
\Delta F_2&=&2\,Q_ea'_eQ_fa'_f\Re\chi\sup_{Z'}
+4\,v'_ea'_ev'_fa'_f|\chi\sup_{Z'}|^2 \nonumber \\[2mm]
& &
+2(v_ea'_e+v'_ea_e)(v_fa'_f+v'_fa_f)\Re(\chi\sup_{Z}\chi^*_{Z'}).
\end{eqnarray}
The coupling constants $v^{(\prime)}_f$, and $a^{(\prime)}_f$ are 
normalized to the unit of charge $e$, and are expressed in terms of the 
couplings in the current basis (\ref{current}) as
\begin{eqnarray}
\label{Eq:couplings}
v_f&=&\frac{g_Z}{e}V^f_Z=\frac{1}{2 s_W c_W}(I^f_{3L}-2Q_f s^2_W),
\qquad
a_f=\frac{g_Z}{e}A^f_Z=\frac{1}{2 s_W c_W}\, I^f_{3L}, \nonumber \\
v'_f&=&\frac{g_{Z'}}{e}V^f_{Z'},
\qquad
a'_f=\frac{g_{Z'}}{e}A^f_{Z'},
\end{eqnarray}
and the gauge boson propagators are
$\chi\sup_V=s/(s-M^2_V+iM_V\Gamma_V)$, $V=Z$, $Z'$.

At $e^+e^-$ energies relevant to our analysis two leptonic 
observables\footnote{We do not
consider observables based on polarization of neither the beams
nor the final fermions, due to their lower sensitivity.}
can be directly measured, the total cross section
\begin{equation}
\sigma_{ff}=\int\limits_{-1}^{1}
\frac{\dd\sigma_{ff}}{\dd\cos\theta}\, \dd\cos\theta
=\sigma_{\rm pt}\, F_1
\end{equation}
with $\sigma_{\rm pt}\equiv\sigma(e^+e^-\to\gamma^\star\to\mu^+\mu^-)=
(4\pi\alpha^2)/(3s)$, and the forward-backward asymmetry defined as
\begin{equation}
\label{Eq:AFB}
A_{\rm FB}=\frac{\int_{0}^{1}(\dd\sigma_{ff}/\dd\cos\theta)\,\dd\cos\theta
-\int_{-1}^{0}(\dd\sigma_{ff}/\dd\cos\theta)\,\dd\cos\theta}
{\int_{0}^{1}(\dd\sigma_{ff}/\dd\cos\theta)\,\dd\cos\theta
+\int_{-1}^{0}(\dd\sigma_{ff}/\dd\cos\theta)\,\dd\cos\theta}=
\frac{3}{4}\frac{F_2}{F_1},
\end{equation}
and obeying $|A_{\rm FB}|\le 3/4$.

The total width $\Gamma_{Z'}$ can be assessed from the
expression
\begin{equation}
\Gamma_{Z'}=\sum_{\rm leptons}\Gamma_{Z'}(Z'\to l\bar l)+
3\sum_{q}\Gamma_{Z'}(Z'\to q\bar q),
\end{equation}
with $\Gamma_{Z'}(Z'\to f\bar f)$ the partial width:
\begin{equation}
\Gamma_{Z'}(Z'\to f\bar f)=\frac{e^2M_{Z'}}{12\pi}
\sqrt{1-4m^2_f/M_{Z'}^2}\left[
{v_f^\prime}^2+{a_f^\prime}^2+2m^2_f/M_{Z'}^2\hskip 2pt
({v_f^\prime}^2-2{a_f^\prime}^2)\right]\, .
\end{equation}

%%%%%%%%%%%%%%%%%%%%%%%%%%%%%%%%%%%%%%%%%%%%%%%%%%%%%%%%%%%%%%%%%%%%%%%%
\section{Interference pattern at the Born level}
%%%%%%%%%%%%%%%%%%%%%%%%%%%%%%%%%%%%%%%%%%%%%%%%%%%%%%%%%%%%%%%%%%%%%%%%
In order to explain qualitatively, in as simple terms as possible,
the interference pattern induced by a $Z'$, we shall first present
Born-level results.
We shall furthermore focus on the leptonic channel (\ref{Eq:leptons})
due to its remarkable phenomenological features as compared with the 
process (\ref{Eq:quarks}). In particular, assuming leptonic universality, 
one avoids further assumptions on the couplings.
In general, there are then only three independent parameters to describe 
the leptonic process, the couplings $v_l'$, $a_l'$ 
and the mass $M_{Z'}$.\footnote{We do not consider here $Z-Z'$ mixing,
because it is important only at $\sqrt{s}=M_Z$, which is outside
the scope of this paper.}

For the purpose of illustrating the structure of the interference
in the observables $\sigma_{\mu\mu}$ and $A_{\rm FB}$,
we choose three representative sets of $Z'$ couplings to leptons with 
fixed mass $M_{Z'}$.
Each choice of $Z'$ parameters is denoted as follows:
vector $Z^\prime_V$ ($v^\prime_l=1$, $a^\prime_l=0$);
axial vector $Z^\prime_A$ ($v^\prime_l=0$, $a^\prime_l=1$);
$Z^\prime$ ($v^\prime_l=1$, $a^\prime_l=1$). (In the SM
$a_l=-0.6$ and $v_l=0.08a_l$ for $s^2_W=0.23$.) 
For all cases we take $M_{Z'}=500$~GeV.
 
It is convenient to present illustrative results
in terms of the observables $\sigma_{\mu\mu}$ and $A_{\rm FB}$,
as well as their deviations from the SM prediction:
\begin{equation}
\frac{\Delta\sigma_{\mu\mu}}{\sigma_{\mu\mu}^{\rm SM}}\equiv
\frac{\sigma_{\mu\mu}-\sigma_{\mu\mu}^{\rm SM}}{\sigma_{\mu\mu}^{\rm SM}}, 
\label{deltasig}
\end{equation}
and
\begin{equation}
\Delta A_{\rm FB}\equiv A_{\rm FB}-A_{\rm FB}^{\rm SM}.
\label{deltaafb}
\end{equation}
Let us consider first the cross section of the process (\ref{Eq:leptons}). 
The energy dependence of the cross section for the three
representative cases of $Z^\prime_V$, $Z^\prime_A$, $Z^\prime$, and the SM
one as well as the deviation (\ref{deltasig}) are shown
in Figs.\ 1a and 1b, respectively. One can understand the $Z'$ 
interference structure using the approximate expression for the relative 
deviation (\ref{deltasig}):
\begin{equation}
\Delta\sigma_{\mu\mu}
\approx\sigma_{\rm pt}
\left[2\,v_l'{}^2\, \Re\,\chi\sup_{Z'}
+2\,a_l^2\,a'_l{}^2\, \Re(\chi\sup_{Z}\chi^*_{Z'})
+(v_l'{}^2+a_l'{}^2)^2|\chi\sup_{Z'}|^2\right],
\label{approx}
\end{equation}
where we take into account the fact that in the SM $|v_l|\ll|a_l|<1$, 
so that one can neglect the terms proportional to $v_l$ 
as compared with those involving $a_l$.

A first observation that can be made from Eq.~(\ref{approx})
is that only {\it squares} of coupling constants appear.
It follows that the signs of the interference terms due to the $Z'$, 
and the deviation $\Delta\sigma_{\mu\mu}$, are governed by 
the propagators of the $Z$ and $Z'$. 
(This is rather unique, as compared with
the process (\ref{Eq:quarks}).)
Furthermore, in certain energy regions, the magnitudes of these 
propagators will be quite different, such that one of
them will dominate.

The first term on the right-hand side of Eq.~(\ref{approx}) describes 
$\gamma-Z'$ interference. It is given by the squared {\it vector} coupling,
and the real part of the $Z'$ propagator.
For a $Z'$ with {\it arbitrary} couplings, we will thus always expect 
destructive $\gamma-Z'$ interference at $\sqrt{s}< M_{Z'}$,
independently of whether $\sqrt{s}$ is below or above $M_Z$.
The second term in Eq.~(\ref{approx}) describes $Z-Z'$ interference.
In contrast to the former, it depends only on the {\it axial} coupling,
$a'_l{}^2$. Since it is proportional to the product of two propagator
factors, it will be {\it positive} below the $Z$, and {\it negative}
above. 
It should be emphasized that all these features are model independent. 
However, the {\it quantitative} interference pattern will depend on
the coupling strengths, and thus be model dependent.

For energies relevant to our analysis we can further simplify
Eq.~(\ref{approx}), neglecting the $Z'$ resonance term as well
as the imaginary part of the $Z'$ propagator,
\begin{eqnarray}
\label{Eq:Delsigma}
\Delta\sigma_{\mu\mu}(Z')
&\underset{\sqrt{s}\ll M_{Z'}}{\approx}&
\Delta\sigma_{\mu\mu}(Z_V') +\Delta\sigma_{\mu\mu}(Z_A') \nonumber \\
&\approx&2\sigma_{\rm pt}\left(v_l'{}^2+a_l'{}^2\,a_l^2\, 
\Re\,\chi\sup_Z\right)\chi\sup_{Z'}\, .
\end{eqnarray}

In order to illustrate the interplay between these interference terms,
let us start from energies attainable at TRISTAN, $\sqrt s<M_Z$.
At TRISTAN, the $\gamma-Z'$ interference terms dominate the $Z'$ 
contributions, and the deviation from the SM is negative due to 
destructive $\gamma-Z'$ interference. 
There are indeed preliminary experimental results showing
such a cross section deficit in the leptonic channel 
\cite{kek2,lep1-rad}.

At higher energies, around the $Z$ boson peak, 
$|M_Z-\sqrt{s}|\approx\Order(\Gamma_Z)$, the $Z-Z'$ interference dominates.
In particular, below the $Z$ resonance the $Z-Z'$ interference 
is constructive and thus tends to compensate the $\gamma-Z'$ 
interference. As can be seen from the approximation (\ref{Eq:Delsigma}),
the vector and axial vector parts are additive.
At an energy $\sqrt{s}=M_Z\sqrt{v_l'{}^2/(v_l'{}^2+a_l^2a_l'{}^2)}$,
where complete compensation of destructive $\gamma-Z'$ and constructive
$Z-Z'$ interferences takes place, 
the deviation $\Delta\sigma_{\mu\mu}$ vanishes.
The point at which this occurs, is independent of the value 
of the $Z'$ mass.
Representative cases are given in Figs.~1a and 1b.
The energy $\sqrt{s}<M_Z$ is suitable for a discrimination
between pure vector and axial vector bosons.
In fact, Fig.~1b shows a negative deviation of the cross section
induced by $Z'_V$, whereas it is positive for $Z'_A$.

On the other hand, at $\sqrt{s}> M_Z$, both interference terms are 
destructive and enhance each other. 
This result is completely independent
of whether the $Z'$ has axial or vector couplings.
Since LEP2 will operate in this region, the effect of a $Z'$
will be to give a cross section {\it deficit} in the leptonic channel.
This is the only region where the effect of a $Z'$ on the cross
section is unambiguous in sign.
In this sense, LEP2 has advantages with respect to TRISTAN and LEP1.

At still higher energies, the pure $Z'$ term in Eq.~(\ref{approx}) 
becomes important and eventually compensates the negative terms.
For some energy, $M_Z<\sqrt{s}<M_{Z'}$,
there is thus another point where complete cancellation occurs,
so that $\Delta\sigma_{\mu\mu}=0$.
These features in the energy dependence of the $Z'$ contributions
can be observed in Figs.~1a and 1b.
In Fig.~1 we took the axial and vector couplings equal, in order
to compare the relative sensitivities to vector and axial couplings,
at different energies.

Another important quantity is the statistical significance, 
\begin{equation}
S[\sigma]=\frac{|\sigma_{\mu\mu}-\sigma_{\mu\mu}^{\rm SM}|}
{\delta\sigma_{\mu\mu}}
=\frac{|\Delta\sigma_{\mu\mu}|}
{\sqrt{\sigma_{\mu\mu}^{\rm SM}}}\,
\sqrt{\Lumint}\, ,
\label{SS}
\end{equation}
where $\delta\sigma_{\mu\mu}$ is the statistical uncertainty,
and $\Lumint$ the integrated luminosity, $\Lumint=\int\dd t{\cal L}$.
It determines the deviation of the cross section from the SM prediction,
in terms of standard deviations.
We show in Fig.~1c the energy dependence of this statistical
significance, for $\Lumint=300~\mbox{pb}^{-1}$, and for two different
masses, $M_{Z'}=500$~GeV and 800~GeV.
As can be seen from the figure, maxima are located at TRISTAN,
LEP1 (``$Z$ line shape") and LEP2 energies.
This feature is rather stable with respect to the $Z'$ mass.
However, the figure also shows that the sensitivity is reduced
at higher $Z'$ masses.

Analogous considerations apply to the forward-backward asymmetry,
Eq.~(\ref{Eq:AFB}).
This asymmetry, as a function of energy, is shown in Fig.~2a.
The deviation from the SM, induced by the $Z'$ interference,
can be expressed as
\begin{equation}
\label{Eq:delAFB}
\Delta A_{\rm FB}
\equiv A_{\rm FB}-A_{\rm FB}^{\rm SM}
\propto \left[
\left(a_l'{}^2-\frac{4}{3}\, A_{\rm FB}^{\rm SM}\, v_l'{}^2\right)
+\left(v_l'{}^2-\frac{4}{3}\, A_{\rm FB}^{\rm SM}\, a_l'{}^2\right)
a_l^2\, \Re\,\chi\sup_Z\right]\chi\sup_{Z'}\, .
\end{equation}
For convenience, it has been split into two terms.

At TRISTAN energies, the main contribution comes from the first term.
Since $A_{\rm FB}^{\rm SM}$ at these energies is small, cf.\ Fig.~2a,
the deviation will essentially be given by the {\it axial} coupling
to the $Z'$. Around the $Z$, the second term becomes important.
Again $A_{\rm FB}^{\rm SM}$ is small, and the main contribution will
be due to the {\it vector} coupling, $v_l'$.
Thus, the forward-backward asymmetry is complementary to the
cross section.
At LEP2, as can be seen from Eq.~(\ref{Eq:delAFB}),
and since $A_{\rm FB}^{\rm SM}$ there is a slowly varying function
of energy, the sensitivities to $a_l'$ and $v_l'$ are comparable.

In Figs.~2b and 2c we show the relative deviation from the SM, 
$\Delta A_{\rm FB}$, and the statistical significance, respectively.
The latter is defined in analogy with Eq.~(\ref{SS}) as
$S[A_{\rm FB}]=|A_{\rm FB}-A_{\rm FB}^{\rm SM}|/
\delta A_{\rm FB}$.
We note from Fig.~2c that the sensitivity $S[A_{\rm FB}]$ has more 
structure than the corresponding one for the cross section.
But still, we can conclude that the energy regions
available at TRISTAN, LEP1 ($\sqrt{s}\ne M_Z$) and LEP2 are 
rather favourable for $Z'$ searches.

%%%%%%%%%%%%%%%%%%%%%%%%%%%%%%%%%%%%%%%%%%%%%%%%%%%%%%%%%%%%%%%%%%%%%%%%
\section{Model independent limits on $Z'$ couplings}
%%%%%%%%%%%%%%%%%%%%%%%%%%%%%%%%%%%%%%%%%%%%%%%%%%%%%%%%%%%%%%%%%%%%%%%%
In writing down the neutral current interaction of the $Z'$ in a
model-independent way we follow \cite{Leike2}, but with different
normalization of the couplings. The $Z'$ mediated amplitude for fermion 
pair production in the Born approximation can be written 
as
\begin{eqnarray}
{\cal M}(Z')&\propto &\frac{g^2_{Z'}}{s-M^2_{Z'}}
\bigl[\bar u_e\gamma_\mu(V^e_{Z'}-A^e_{Z'}\gamma_5)u_e\bigr]
\bigl[\bar u_f\gamma^\mu(V^f_{Z'}-A^f_{Z'}\gamma_5)u_f\bigr] \nonumber \\
&=&-\frac{4\pi}{M^2_Z}
\left[\bar u_e\gamma_\mu(V_e-A_e\gamma_5)u_e\right]
\left[\bar u_f\gamma^\mu(V_f-A_f\gamma_5)u_f\right], 
\end{eqnarray}
with 
\begin{equation}
\label{coupl}
V_f=V^f_{Z'}\, \sqrt{\frac{g^2_{Z'}}{4\pi}\,
\frac{M^2_Z}{M^2_{Z'}-s}}, 
\qquad
A_f=A^f_{Z'}\, \sqrt{\frac{g^2_{Z'}}{4\pi}\,
\frac{M^2_Z}{M^2_{Z'}-s}}.
\end{equation}

It should be noted that the imaginary part of 
the $Z'$ propagator is irrelevant up to LEP2 energies,
hence it is set to zero in the present analysis. 
In the parameterization (\ref{coupl}), the lepton pair production
via $Z'$ boson exchange can be described by effective four-fermion contact 
interactions \cite{Leike2,Eichten}.

The sensitivity of observables, e.g., of the total cross section 
$\sigma_{\mu\mu}$, has been assessed numerically by defining
a $\chi^2$ function as follows:
\begin{equation}
\label{Eq:chisq}
\chi^2
=\left(\frac{\Delta\sigma_{\mu\mu}}{\delta\sigma_{\mu\mu}}\right)^2,
\end{equation}
where $\Delta\sigma_{\mu\mu}$ is given by Eq.~(\ref{deltasig}) and
the uncertainty $\delta\sigma_{\mu\mu}$ combines both statistical 
and systematic errors. As a criterion to derive allowed regions for
the coupling constants if no deviations
from the SM were observed, and in this way to assess the sensitivity
of the process (\ref{Eq:leptons}) to $V_l$ and $A_l$,
we impose that $\chi^2<\chi^2_{\rm crit}$, where $\chi^2_{\rm crit}$
is a number that specifies the desired level of significance.

We start our discussion from the constraints on the $Z'$ lepton couplings 
at Born level, and in the sequel take into account the radiative 
corrections. 
In this analysis we will take the achieved luminosities relevant to
the leptonic process (\ref{Eq:leptons}) at TRISTAN, LEP1, and LEP1.5,
summarized in Table~1.
Also, we will use the expected luminosity at the future experiments 
at LEP2 in order to obtain the allowed regions of the couplings 
$V_l$ and $A_l$.
Statistical and systematic errors are added in quadrature.

\begin{table}
\begin{center}
\begin{tabular}{||l|r|r|r||}                     \hline
           &$\sqrt{s}$ [GeV] & $\delta_{\rm syst}$ [\%] & 
$\Lumint$ or $N_{\mu\mu}$ \\ \hline
TRISTAN \cite{kek2}     & 58 & $1$ & 300~pb${}^{-1}$ \\
LEP1 \cite{lep1}       & $M_Z\mp 2$  & $0.3$ &3299/5505 \\ 
LEP1.5 \cite{lep1_5}   & 130--140 & $1$   & 5~pb${}^{-1}$\\ 
LEP2 \cite{lep2}       & 190      & $1$   & 500~pb${}^{-1}$ \\ \hline
\end{tabular}
\end{center}
\caption{Systematic errors and integrated luminosities
(or number of leptonic pair production events) available at $e^+e^-$ 
colliders. The parameters correspond to a typical experiment.}
\end{table}

According to Eqs.~(\ref{Eq:Delsigma}), (\ref{coupl}) and (\ref{Eq:chisq}) 
one can conclude that the cross section $\sigma_{\mu\mu}$ yields 
ranges of observability in the ($A_l,\ V_l$) plane bounded either by 
hyperbolas at $\sqrt{s}<M_Z$, or by ellipses around the origin 
for $\sqrt{s}>M_Z$. In contrast to the cross section, 
the forward-backward asymmetry $A_{\rm FB}$, as can be seen from 
Eqs.~(\ref{Eq:delAFB}), (\ref{coupl})
and (\ref{Eq:chisq}), yields the detectability region as bounded by
an ellipse at the TRISTAN energy, and by hyperbolas at the energy available 
at LEP2.

As an illustration, we show in
Fig.~3 the allowed regions at the level of two standard deviations 
($\chi^2_{\rm crit}=4$) for the $Z'$ couplings to leptons, $V_l$ and $A_l$. 
These exclusion regions, valid for $\sqrt{s}=58$~GeV, are indeed bounded 
by hyperbolas and ellipses, 
as expected from the crude analysis based on Eqs.~(\ref{Eq:Delsigma}) 
and (\ref{Eq:delAFB}).
One can see that these two observables are complementary in obtaining 
bounds on the leptonic couplings $V_l$ and $A_l$. 

%%%%%%%%%%%%%%%%%%%%%%%%%%%%%%%%%%%%%%%%%%%%%%%%%%%%%%%%%%%%%%%%%%%%%%%%
\section{Quantitative results and concluding remarks}
%%%%%%%%%%%%%%%%%%%%%%%%%%%%%%%%%%%%%%%%%%%%%%%%%%%%%%%%%%%%%%%%%%%%%%%%
Up to now we studied the indirect (propagator) effects of new neutral
bosons on the observables $\sigma_{\mu\mu}$ and $A_{\rm FB}$ in the Born 
approximation, ignoring radiative corrections. We note, however, 
that to the extent that the deviations from the SM predictions are small, 
which is indeed the case in our analysis, we can directly confront our 
observables in the Born approximation with the radiatively corrected data. 
While the contribution of the weak loop corrections will only change 
the overall normalization, the photonic corrections and in particular 
those due to initial state radiation give very large contributions 
which depend on the experimental set-up.
Due to the radiative return to the $Z$ resonance at $\sqrt{s}>M_Z$ the 
energy spectrum of the radiated photons is peaked around 
$E_\gamma/E_{\rm beam}\approx 1-M^2_Z/s$ \cite{Djouadi}.
In order to increase the $Z'$ signal,
events with hard photons should be eliminated from a $Z'$ search by a cut 
on the photon energy, $\Delta=E_\gamma/E_{\rm beam}< 1-M^2_Z/s$.

A numerical analysis has been performed by means of the program
ZEFIT \cite{zefit}, which has to be used along with ZFITTER 
\cite{zfitter}.
In this way, all the SM corrections, as well as those of QED 
associated with the $Z'$ contributions were taken into account.
In Fig.~4 we compare the discovery potentials of different accelerators,
with inputs from Table~1. The contours are obtained from a combination
of both observables, $\sigma_{\mu\mu}$ and $A_{\rm FB}$,
and correspond to two standard deviations.
From the currently available data 
one sees that TRISTAN provides the better sensitivity.
Furthermore, the figure includes two contours for LEP2,
corresponding to 150~pb${}^{-1}$ (one year of running) and
500~pb${}^{-1}$ (three years of running).
After one year's operation, LEP2 will achieve a higher
sensitivity than TRISTAN.

Finally, we discuss how the bounds on the couplings $V_l$ and $A_l$
depend on energy and luminosity.
This is a very important aspect of the evaluation of the physics
potential of the colliders and particularly interesting when one wants
to compare LEP2 with TRISTAN.
For this purpose, let us now reverse our attitude, and assume that 
the two-sigma deficit of the leptonic cross section seen 
at TRISTAN \cite{kek2} represents a real effect.
What is then the luminosity required at LEP2 to confirm this effect
at sufficient significance (e.g., five standard deviations)?
This question can be answered from the approximate scaling law
\begin{equation}
\Lumint(b)
=\frac{\chi^2_{\rm crit}(b)}{\chi^2_{\rm crit}(a)}\,
\frac{s(a)}{s(b)}\, \Lumint(a)
\end{equation}
which provides 170~pb${}^{-1}$ for a TRISTAN luminosity of
300~pb${}^{-1}$. This relation follows from the Born-level
formulas for the cross section, neglecting systematic errors.
Allowing for radiative corrections and systematic errors,
we find that the luminosity required is 100--120~pb${}^{-1}$.
This reduction of luminosity is mainly due to the systematic uncertainty.
Thus, after less than one year of operation, LEP2 will be able to
achieve a sensitivity which allows for a check of the preliminary
TRISTAN results \cite{kek2} with a significance better than $5\sigma$.

In conclusion, we have reviewed the interference effects induced by
an extra neutral gauge boson $Z'$ in the process 
$e^+e^-\to\mu^+\mu^-$.
Assuming lepton universality, the lepton channel has the advantage 
over the $q\bar q$ channel
that the signs of the interference terms are given very simply by the
propagators of $Z$ and $Z'$. This is caused by the fact that 
the observables under consideration, $\sigma_{\mu\mu}$ and $A_{\rm FB}$, 
depend only on {\it squares} of coupling constants.

At LEP2 energies, the effect of a $Z'$ (with arbitrary
vector and axial vector couplings) is to give a cross section deficit,
$\Delta\sigma_{\mu\mu}<0$, as compared with the SM expectation.
This unique property of $\sigma_{\mu\mu}$ is due to the fact that
the $\gamma-Z'$ as well as the $Z-Z'$ interference terms are
both negative.

At TRISTAN and LEP1 (``$Z$ line shape'') energies the situation is less
definite, due to the interplay between the $\gamma-Z'$ and $Z-Z'$ 
interferences. However, at TRISTAN energies, the cross section is more
sensitive to vector couplings, $V_l$, whereas at LEP1 there is greater
sensitivity to axial couplings, $A_l$.
This is mostly due to the smallness of the vector coupling of the
standard $Z$ to leptons, $|v_l|\ll|a_l|$.
For the forward-backward asymmetry, the situation is reversed, at
TRISTAN there is more sensitivity to $A_l$, whereas around the
$Z$ boson, there is more sensitivity to $V_l$. Thus, $A_{\rm FB}$
is complementary to $\sigma_{\mu\mu}$.

The comparison of the $Z'$ search potential of the $e^+e^-$ colliders
TRISTAN, LEP1 and LEP2 shows the advantage of the latter one.
In particular, after one year of operation, with accumulated luminosity
$\sim150$~pb${}^{-1}$, LEP2 will achieve a $Z'$ sensitivity that
exceeds the current one at TRISTAN.
\medskip

%%%%%%%%%%%%%%%%%%%%%%%%%%%%%%%%%%%%%%%%%%%%%%%%%%%%%%%%%%%%%%%%%%%%%%%%
%\section*{Acknowledgements}

We would like to express our gratitude to Dr.\ A. Babich for the help
with numerical calculations presented in Fig.~4.
This research has been supported by the Research Council of Norway.
%%%%%%%%%%%%%%%%%%%%%%%%%%%%%%%%%%%%%%%%%%%%%%%%%%%%%%%%%%%%%%%%%%%%%%%%

%%%%%%%%%%%%%%%%%%%%%%%%%%%%%%%%%%%%%%%%%%%%%%%%%%%%%%%%%%%%%%%%%%%%%%%%

%\end{document}       % These two lines should be commented
%\endinput            % when running with figures

%%%%%%%%%%%%%%%%%%%%%%%%%%%%%%%%%%%%%%%%%%%%%%%%%%%%%%%%%%%%%%%%%%%%%%%%

\setcounter{figure}{0}
\begin{figure}
\begin{center}
\setlength{\unitlength}{1cm}
\begin{picture}(14.0,14.0)
\put(-1.,0.0){
\mbox{\epsfysize=12cm\epsffile{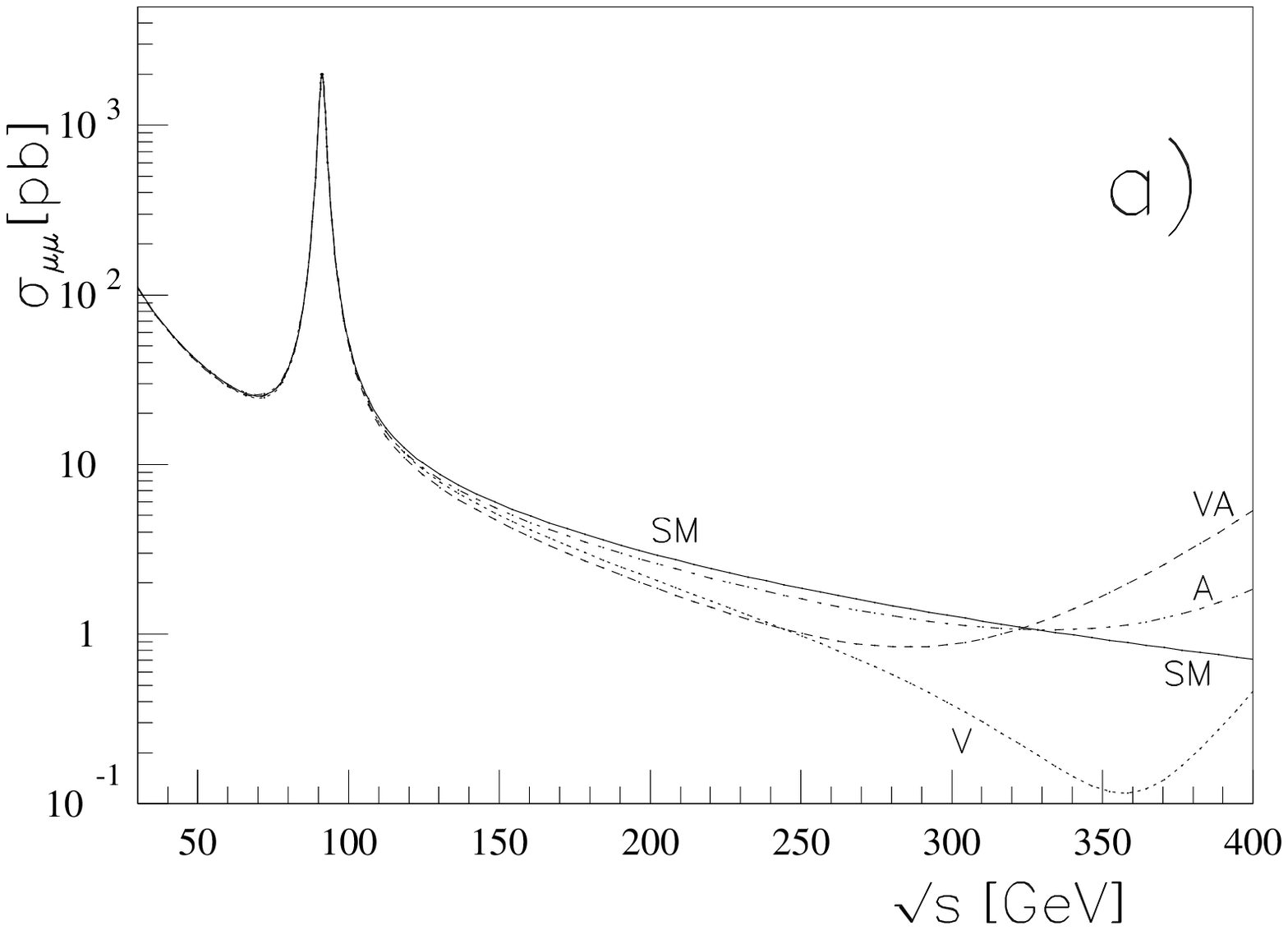}}}
\end{picture}
\caption{
(a) Total cross section for muon pair production $\sigma_{\mu\mu}$
in the Born approximation {\it vs.} c.m.\ energy in the SM and
in the presence of a $Z^\prime$ with mass $M_{Z^\prime}=500$~GeV for the
following cases:
vector (V) $Z^\prime_{\rm V}$ ($v^\prime_l=1$, $a^\prime_l=0$),
axial vector (A) $Z^\prime_{\rm A}$ ($v^\prime_l=0$, $a^\prime_l=1$)
and mixed (VA) $Z^\prime$  ($v^\prime_l=1$, $a^\prime_l=1$).}
\end{center}
\end{figure}
\setcounter{figure}{0}
%%%%%%%%%%%%%%%%%%%%%%%%%%%%%%%%%%%%%%%%%%%%%%%%%%%%%%%%%%%%%%%%%%%%%%%%
%\clearpage

\begin{figure}
\begin{center}
\setlength{\unitlength}{1cm}
\begin{picture}(14.0,14.0)
\put(-1.,0.0){
\mbox{\epsfysize=12cm\epsffile{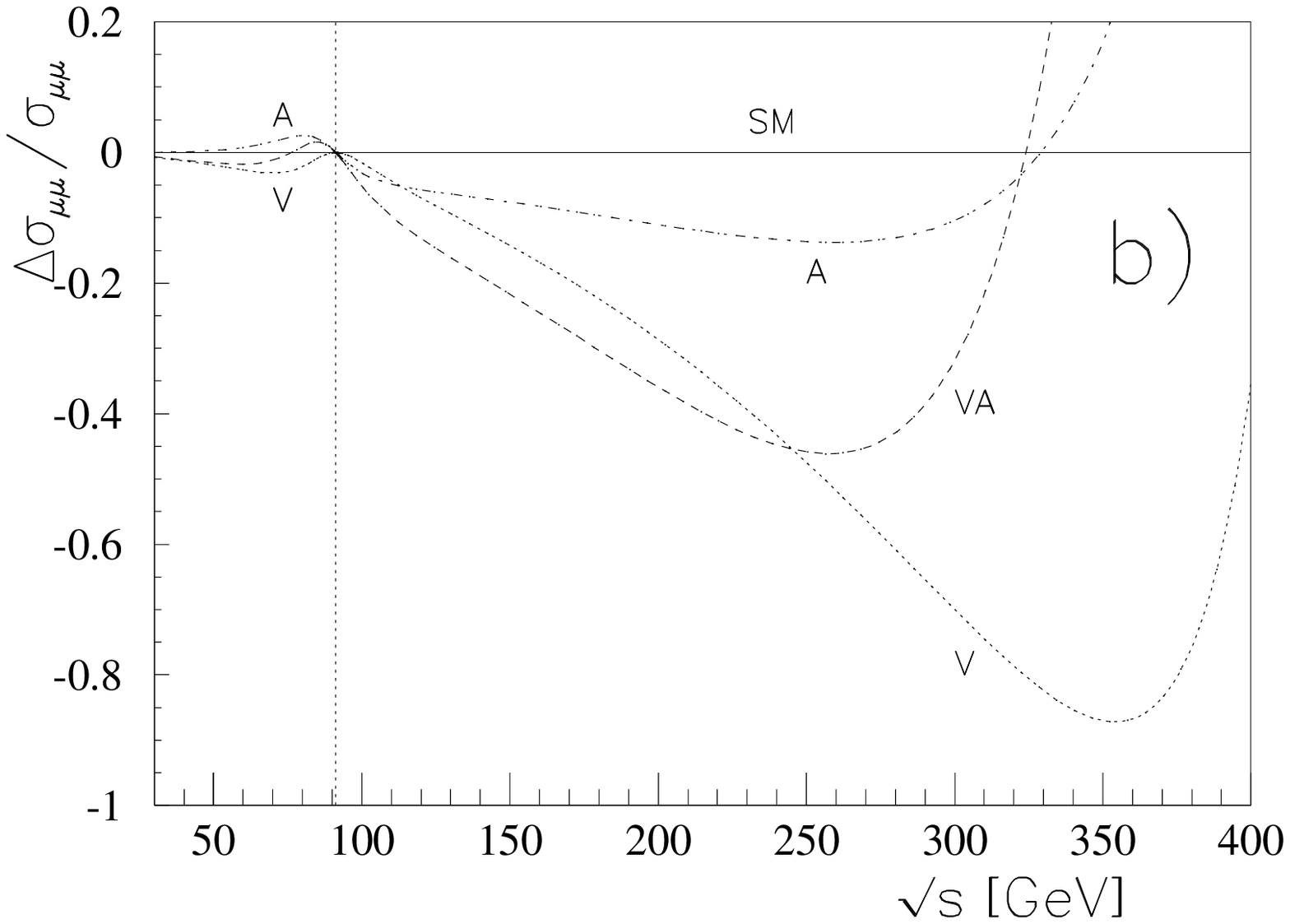}}}
\end{picture}
\caption{
(b) As in (a), but for relative deviation of the cross section,
Eq.~(\ref{deltasig}).}
\end{center}
\end{figure}
\setcounter{figure}{0}
%%%%%%%%%%%%%%%%%%%%%%%%%%%%%%%%%%%%%%%%%%%%%%%%%%%%%%%%%%%%%%%%%%%%%%%%
%\clearpage

\begin{figure}
\begin{center}
\setlength{\unitlength}{1cm}
\begin{picture}(14.0,14.0)
\put(-1.,0.0){
\mbox{\epsfysize=12cm\epsffile{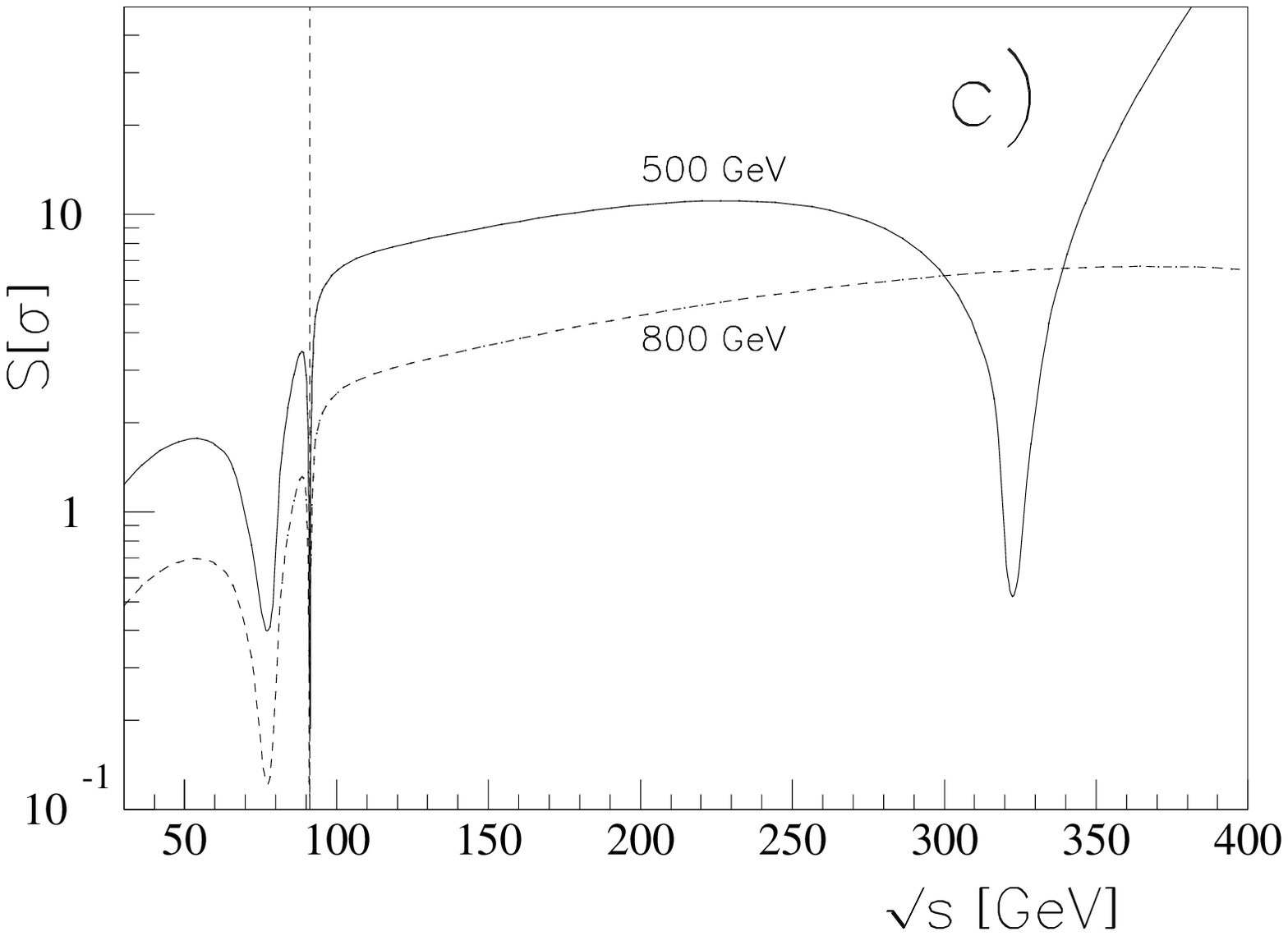}}}
\end{picture}
\caption{
(c) As in (a), but for the statistical significance $S[\sigma]$ 
of Eq.~(\ref{SS}) at $M_{Z'}=500$~GeV and $M_{Z'}=800$~GeV, 
and for integrated luminosity $\Lumint=300~\mbox{pb}^{-1}$.}
\end{center}
\end{figure}
%%%%%%%%%%%%%%%%%%%%%%%%%%%%%%%%%%%%%%%%%%%%%%%%%%%%%%%%%%%%%%%%%%%%%%%%
%\clearpage

\begin{figure}
\begin{center}
\setlength{\unitlength}{1cm}
\begin{picture}(14.0,14.0)
\put(-1.,0.0){
\mbox{\epsfysize=12cm\epsffile{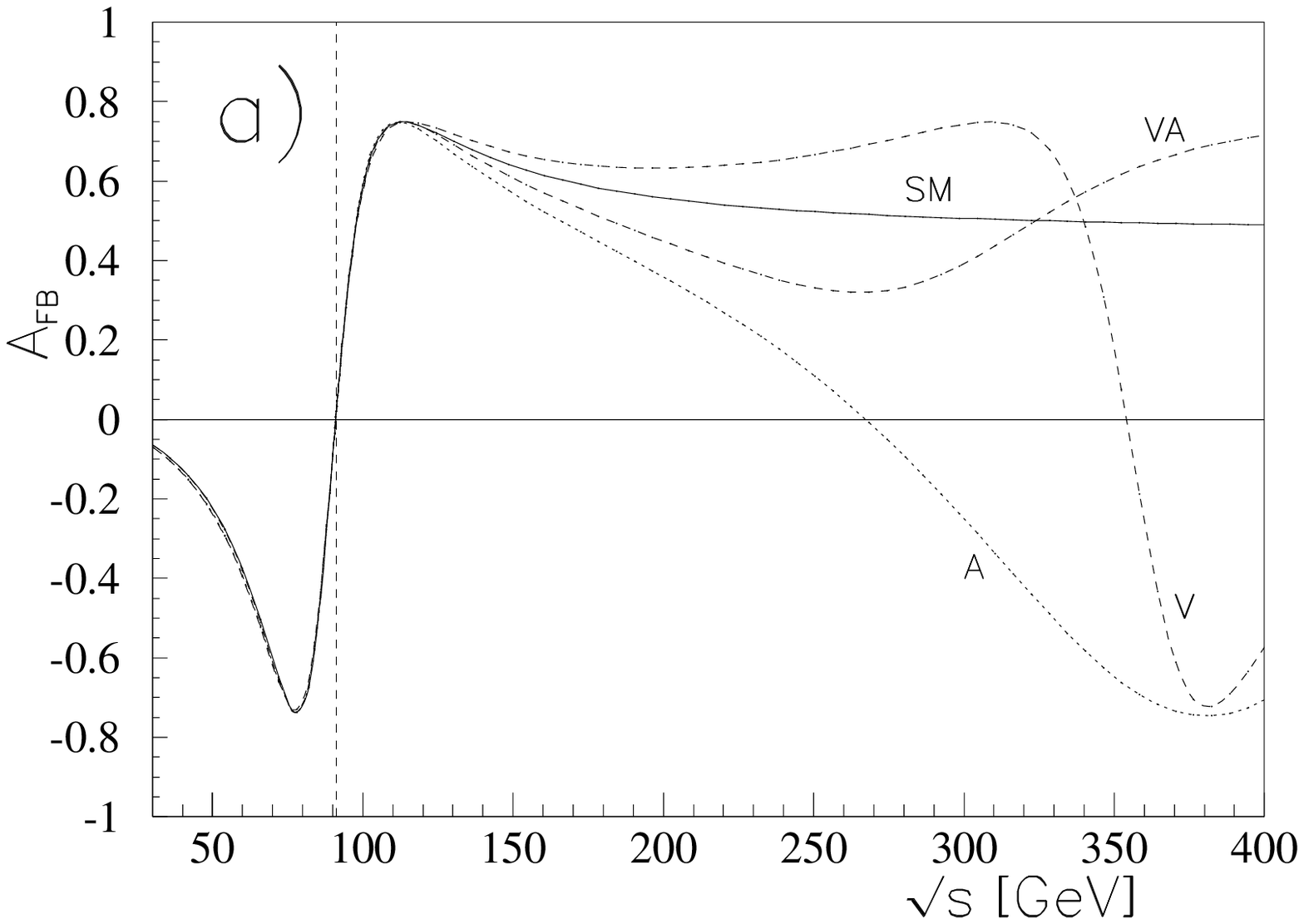}}}
\end{picture}
\caption{
(a) As in Fig.~1, but for the forward-backward asymmetry $A_{FB}$.}
\end{center}
\end{figure}
\setcounter{figure}{1}
%%%%%%%%%%%%%%%%%%%%%%%%%%%%%%%%%%%%%%%%%%%%%%%%%%%%%%%%%%%%%%%%%%%%%%%%
%\clearpage

\begin{figure}
\begin{center}
\setlength{\unitlength}{1cm}
\begin{picture}(14.0,14.0)
\put(-1.,0.0){
\mbox{\epsfysize=12cm\epsffile{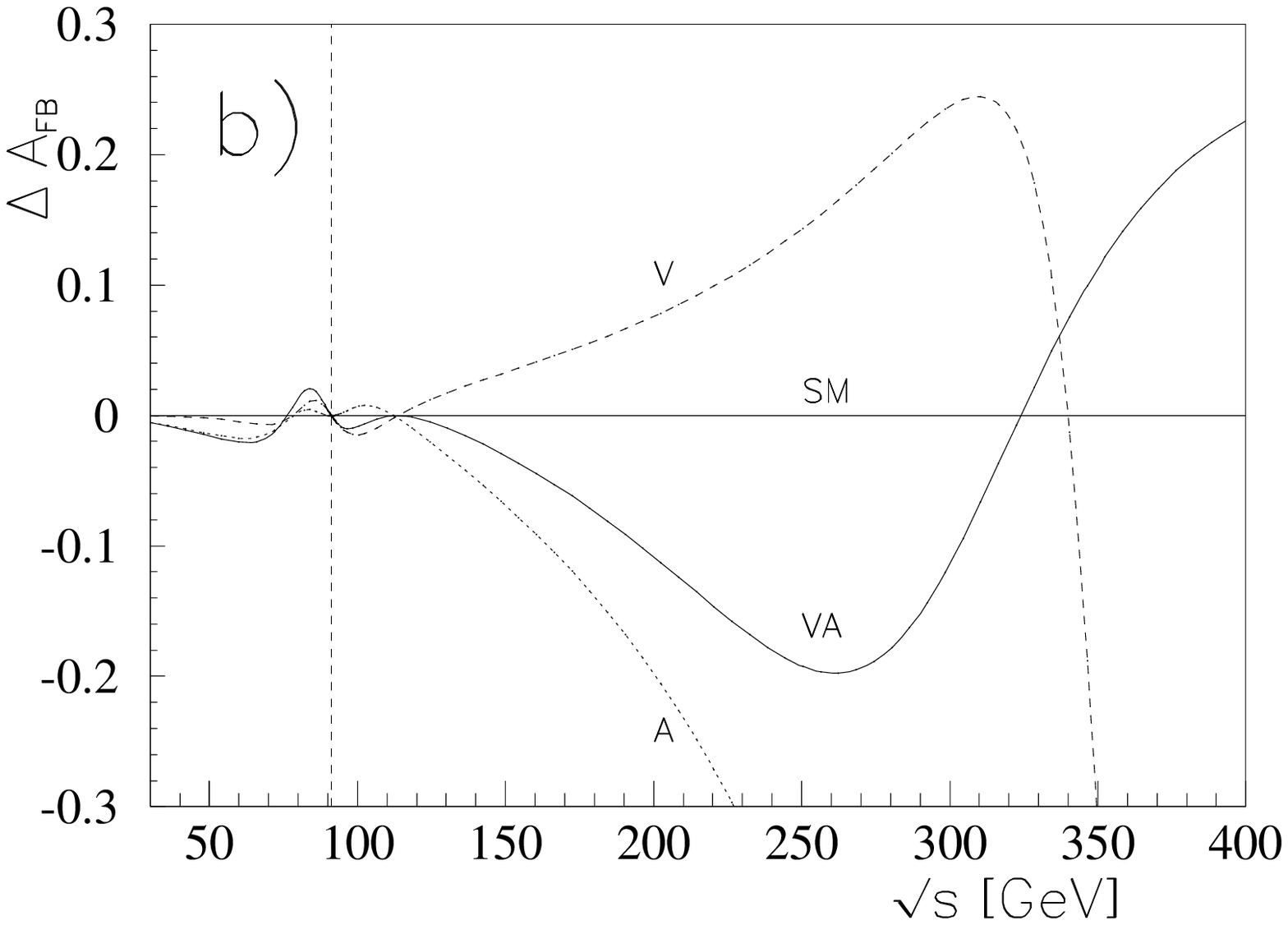}}}
\end{picture}
\caption{
(b) As in Fig.~1, but for the forward-backward asymmetry $A_{FB}$.}
\end{center}
\end{figure}
\setcounter{figure}{1}
%%%%%%%%%%%%%%%%%%%%%%%%%%%%%%%%%%%%%%%%%%%%%%%%%%%%%%%%%%%%%%%%%%%%%%%%
%\clearpage

\begin{figure}
\begin{center}
\setlength{\unitlength}{1cm}
\begin{picture}(14.0,14.0)
\put(-1.,0.0){
\mbox{\epsfysize=12cm\epsffile{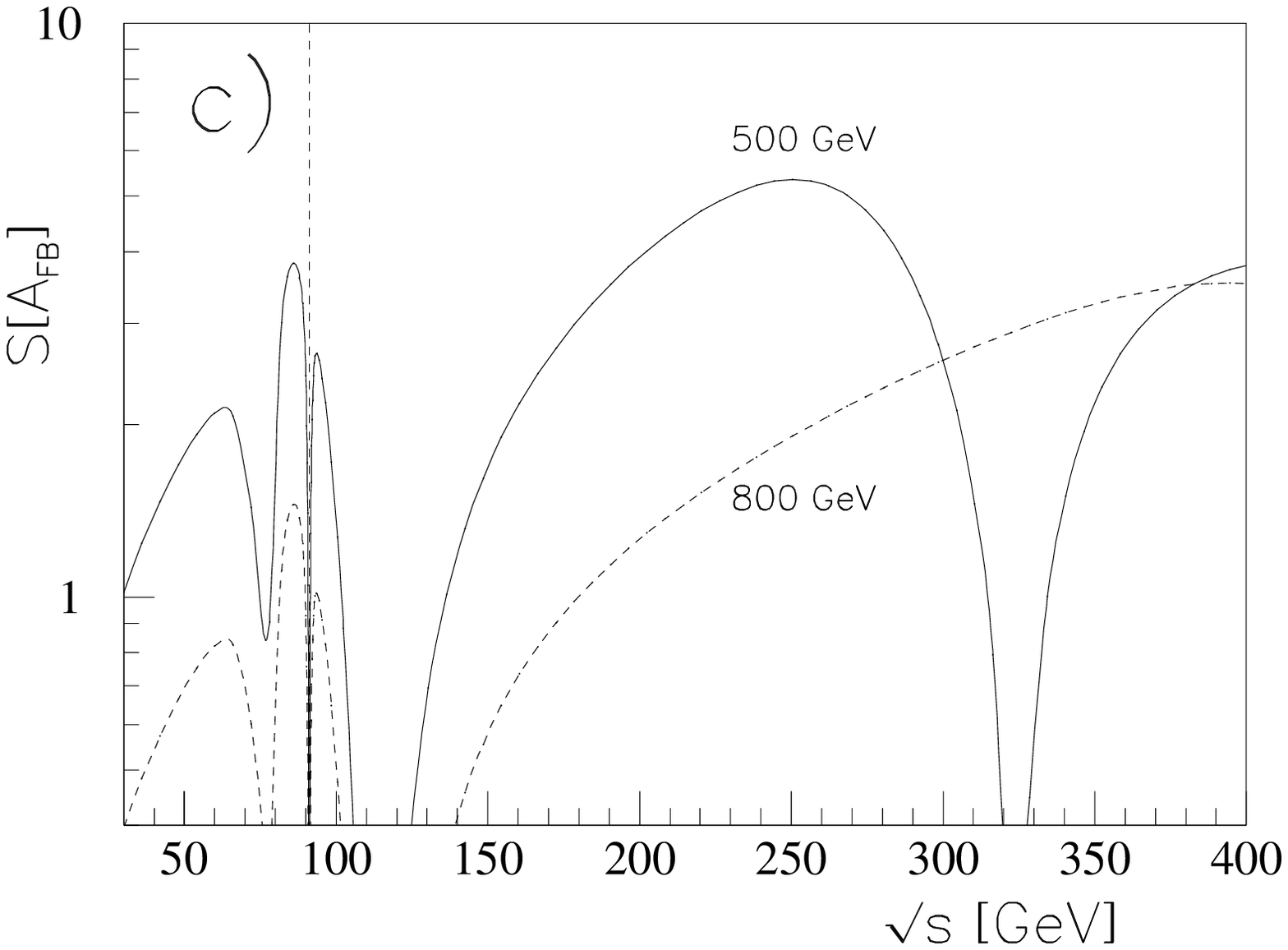}}}
\end{picture}
\caption{
(c) As in Fig.~1, but for the forward-backward asymmetry $A_{FB}$.}
\end{center}
\end{figure}                  
%%%%%%%%%%%%%%%%%%%%%%%%%%%%%%%%%%%%%%%%%%%%%%%%%%%%%%%%%%%%%%%%%%%%%%%%
%\clearpage

\begin{figure}
\begin{center}
\setlength{\unitlength}{1cm}
\begin{picture}(16.0,16.0)
\put(1.,0.0){
\mbox{\epsfysize=14cm\epsffile{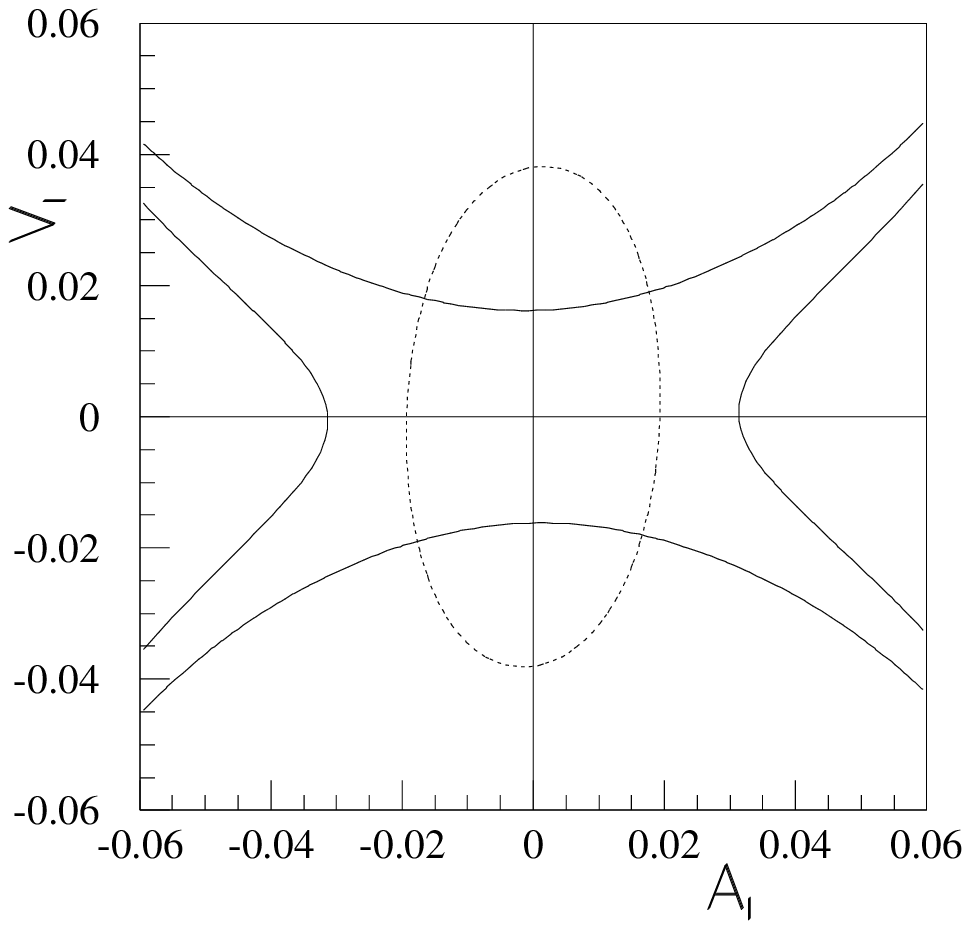}}}
\end{picture}
\caption{
Upper bounds (in the Born approximation) for model-independent
couplings ($A_l$, $V_l$) at the level of two standard deviations,
derived from $\sigma_{\mu\mu}$ (hyperbolas) and $A_{\rm FB}$
(ellipse) at TRISTAN.}
\end{center}
\end{figure}

%%%%%%%%%%%%%%%%%%%%%%%%%%%%%%%%%%%%%%%%%%%%%%%%%%%%%%%%%%%%%%%%%%%%%%%%
%\clearpage

\begin{figure}
\begin{center}
\setlength{\unitlength}{1cm}
\begin{picture}(16.0,16.0)
\put(1.,0.0){
\mbox{\epsfysize=14cm\epsffile{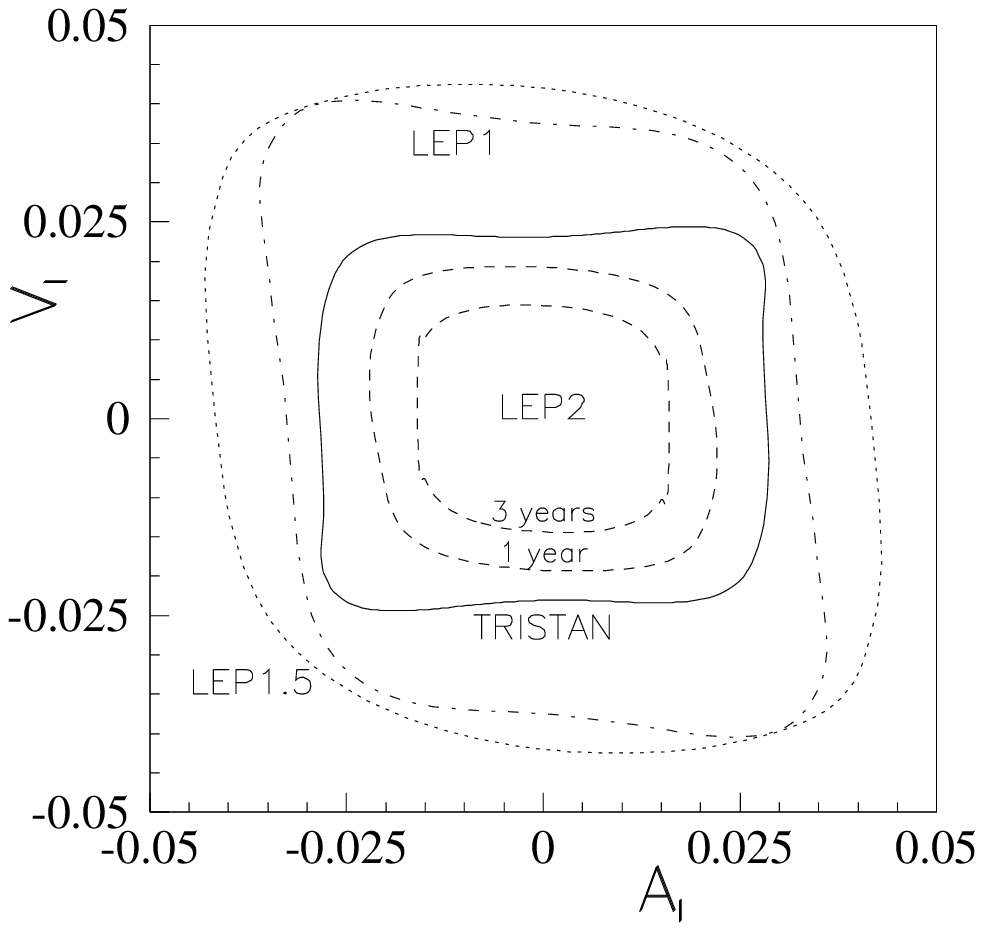}}}
\end{picture}
\caption{
Combined regions allowed by $\sigma_{\mu\mu}$ and $A_{\rm FB}$ 
at the $2\sigma$ level in the ($A_l$, $V_l$) plane 
for TRISTAN, LEP1, LEP1.5 and LEP2 colliders, 
for inputs given in Table 1. 
Two bounds are shown for LEP2 corresponding to 
$\Lumint=150~\mbox{pb}^{-1}$ (one year of running) and 
$\Lumint=500~\mbox{pb}^{-1}$ (three years of running).
Radiative corrections are included.}
\end{center}
\end{figure}              


\newpage
\begin{thebibliography}{99}

\bibitem{Langacker}
Precision Tests of the Standard Electroweak Model,
{\it Advanced Series on Directions in High Energy Physics}, Vol.14, ed.
P. Langacker, World Scientific, 1995.

\bibitem{Warsaw}
A. Blondel, talk at the International Conference on High Energy Physics,
Warsaw, Poland, July 1996.

% Superstring-inspired E_6 models:
\bibitem{Rizzo} For a review see, e.g., J. L. Hewett and T. G. Rizzo, 
Physics Reports {\bf 183} (1989) 193.

% Discovery limits for future colliders:
\bibitem{Godfrey} S. Godfrey, Phys. Rev. D {\bf 51} (1995) 1402.

\bibitem{Maeshima}
K. Maeshima (CDF Collaboration),
Presented at 28th International Conference on High-energy
Physics (ICHEP 96), 
Warsaw, Poland, 25-31 Jul 1996; 
FERMILAB-CONF-96-412-E.

\bibitem{lep2}
Physics at LEP2, Proceedings of the Workshop, Geneva, Switzerland (1996), 
CERN 06-01, G. Altarelli, T. Sj{\"o}strand and F. Zwirner, eds.

\bibitem{Leike1} A. Leike and S. Riemann, report DESY 96-111 (1996).

\bibitem{Leike2} A. Leike, Z. Phys. {\bf C62} (1994) 265.

\bibitem{kek1}
Proceedings of the Workshop on
TRISTAN Physics at High Luminosities, Ed. M. Yamauchi,
KEK Proceedings 93-2, April 1993.\\
Proceedings of the 2nd Workshop on
TRISTAN Physics at High Luminosities, Eds H. Sagawa et al.,
KEK Proceedings 93-22, March 1994.\\
M. Sakuda, Nuovo Cim. {107A} (1994) 2389.

\bibitem{kek2} K. Miyabayashi, Recent electroweak results from TRISTAN,
talk presented at ``Mo\-riond-95''.

\bibitem{lep1-rad}
P. Acton et al., Phys.\ Lett.\ {\bf B273} (1991) 338; \\
F. Cao et al., DELPHI 96-80 CONF 12 (1996); \\
M. Acciarri et al., Phys.\ Lett.\ {\bf B374} (1996) 331; \\
ALEPH Collaboration, Contribution to the 28th
International Conference on High Energy Physics, 
Warsaw, Poland, July 1996, PA-07-070.

% Extra Z' gauge boson at TRISTAN:
\bibitem{p1}
A.A. Pankov and I.S. Satsunkevich, Sov. J. Nucl. Phys. {\bf 47} (1988) 849;
Nuovo Cimento  {\bf 103A} (1990) 1121.\\
A.A. Pankov and C. Verzegnassi, Phys. Lett. {\bf B233} (1989) 259.

% Isoscalar bosons:
\bibitem{p2}
A.A. Likhoded, A.A. Pankov and O.P. Yushchenko,
Int. J. Mod. Phys., {\bf A7, v.22} (1992) 1537.\\
S.S. Gershtein, A.A. Likhoded, A.A. Pankov and O.P. Yushchenko,
Phys. Lett. {\bf275B} (1992) 169; Z. Phys. {\bf C56} (1992) 279.
  
\bibitem{kuroda}
M. Kuroda, D. Schildknecht and
K.-H. Schwarzer, Nucl. Phys. {\bf B261} (1985) 432;\\
M. Kuroda, F.M. Renard and D. Schildknecht, Phys. Lett. {\bf B183} (1987) 
366.

% Contact interactions:
\bibitem{Eichten}
E.J. Eichten, K.D. Lane and M.E. Peskin, Phys.\ Rev.\ Lett.\ {\bf 50} (1983) 
811.

\bibitem{lep1}
ALEPH Collaboration, Contribution to the 28th
International Conference on High Energy Physics, 
Warsaw, Poland, July 1996, PA-07-069.

\bibitem{lep1_5}
L3 collaboration, M. Acciari et al., preprint CERN-PPE/95-191, CERN, 
1995;\\
L3 collaboration, Contribution to the 28th
International Conference on High Energy Physics, 
Warsaw, Poland, July 1996, PA-07-042.

% New gauge bosons:
\bibitem{Djouadi}
A. Djouadi, A. Leike, T. Riemann, D. Schaile and C. Verzegnassi,
Z. Phys. {\bf C56} (1992) 289.

\bibitem{zefit}
S. Riemann, FORTRAN program ZEFIT Version 4.2.

\bibitem{zfitter}
D. Bardin et al., preprint CERN-TH. 6443/92, CERN, 1992.


\end{thebibliography}
\end{document}